\documentclass[aps,prl,superscriptaddress,twocolumn]{revtex4} 
\usepackage{graphicx}
\begin{document}

\title{Constraints on new physics from the quark mixing unitarity
    triangle}

\collaboration{\textbf{UT}\textit{fit} Collaboration}
%\homepage{http://www.utfit.org} 
\noaffiliation
\author{M.~Bona}
\affiliation{Laboratoire d'Annecy-le-Vieux de Physique des Particules
LAPP, IN2P3/CNRS, Universit{\'e} de Savoie}
\author{M.~Ciuchini}
\affiliation{Dip. di Fisica, Universit{\`a} di Roma Tre
      and INFN,  Sez. di Roma Tre, I-00146 Roma, Italy}
\author{E.~Franco}
\affiliation{Dip. di Fisica, Universit\`a di Roma ``La Sapienza'' and
  INFN, Sez. di Roma, I-00185 Roma, Italy}
\author{V.~Lubicz}
\affiliation{Dip. di Fisica, Universit{\`a} di Roma Tre
      and INFN,  Sez. di Roma Tre, I-00146 Roma, Italy}
\author{G.~Martinelli}
\affiliation{Dip. di Fisica, Universit\`a di Roma ``La Sapienza'' and
  INFN, Sez. di Roma, I-00185 Roma, Italy}
\author{F.~Parodi}
\affiliation{ Dip. di Fisica, Universit\`a di Genova and INFN, I-16146
  Genova, Italy} 
\author{M.~Pierini}
\affiliation{Department of Physics, University of Wisconsin, Madison,
  WI 53706, USA }
\author{P.~Roudeau}
\affiliation{Laboratoire de l'Acc\'el\'erateur Lin\'eaire, IN2P3-CNRS et
  Universit\'e de Paris-Sud, BP 34, 
      F-91898 Orsay Cedex, France}
\author{C.~Schiavi}
\affiliation{ Dip. di Fisica, Universit\`a di Genova and INFN, I-16146
  Genova, Italy} 
\author{L.~Silvestrini}
\affiliation{Dip. di Fisica, Universit\`a di Roma ``La Sapienza'' and
  INFN, Sez. di Roma, I-00185 Roma, Italy}
\author{A.~Stocchi}
\affiliation{Laboratoire de l'Acc\'el\'erateur Lin\'eaire, IN2P3-CNRS et
  Universit\'e de Paris-Sud, BP 34, 
      F-91898 Orsay Cedex, France}
\author{V.~Vagnoni}
\affiliation{INFN, Sez. di Bologna,  I-40126 Bologna, Italy}

\begin{abstract}
  The status of the Unitarity Triangle beyond the Standard Model
  including the most recent results on $\Delta m_s$, on dilepton
  asymmetries and on width differences is presented. Even allowing for
  general New Physics loop contributions the Unitarity
  Triangle must be very close to the Standard Model result. 
  With the new measurements from the Tevatron, we obtain for
  the first time a significant constraint on New Physics in the
  $B_s$ sector. We present the allowed ranges of New Physics
  contributions to $\Delta F=2$ processes, and of the time-dependent
  CP asymmetry in $B_s \to J/\Psi \phi$ decays.
\end{abstract}

\maketitle

In the last decade, flavour physics has witnessed unprecedented
experimental and theoretical progress, opening the era of precision
flavour tests of the Standard Model (SM). The advent of $B$ factories,
with the measurements of the angles of the Unitarity Triangle (UT),
has opened up the possibility of the simultaneous determination of
SM and New Physics (NP) parameters in the flavour sector. As shown
below, with the most recent improvements obtained at the $B$ factories
and at the Tevatron, the UT analysis in the presence of NP has reached an
accuracy comparable to the SM analysis, providing at the same time
very stringent constraints on NP contributions to $\Delta F=2$
processes.

While in general all the constraints have been improved, three remarkable results have boosted the precision of the UT analysis beyond the SM. First, the CDF
collaboration presented the first measurement of $B_s$--$\bar B_s$
mass difference $\Delta m_{s}$~\cite{dmsCDF}, which reduces the
uncertainty of the SM fit~\cite{instantSM} and has a strong impact on
the determination of the Universal Unitarity Triangle (UUT) \cite{uut}
in models with Minimal Flavour Violation (MFV)
\cite{mfv,gino}. Moreover, it allows for the first
time to put a bound on the absolute value of the amplitude for $B_s$
oscillations~\cite{npcbs}.  Second, the measurement of the dimuon
asymmetry in $p\bar p$ collisions by the D0 experiment~\cite{dimuonD0}
can be translated into a bound on the phase of the same amplitude~\cite{nir}. Third, the
measurements of the width difference for $B_q$ mesons provide another
constraint on the phase of the mixing amplitudes, complementary to the
one given by dilepton asymmetries~\cite{dgammanp}.

In this Letter, we first discuss extensions of the SM with
MFV, in which no new source of flavour and CP violation is present
beyond the SM Yukawa couplings. We analyze the impact of $\Delta
m_{s}$ on the UUT determination, where the ratio $\Delta m_{d}/\Delta
m_{s}$ plays a crucial role since it is independent of NP
contributions. We find that the UUT analysis has now an accuracy very
close to the SM UT fit. Using instead the information coming from the
individual measurements of $\Delta m_{s}$, $\Delta m_{d}$ and
$\varepsilon_K$, we constrain NP contributions to the $\Delta F=2$
Hamiltonian, both in the small and large $\tan \beta$ regimes. We find
improved constraints on the NP scale $\Lambda$ that suppresses
non-renormalizable effective interactions. 

We then turn to the most general case in which NP contributions with
an arbitrary phase are allowed in all
sectors, and obtain a fully model-independent determination of the CKM
parameters $\bar \rho$ and $\bar \eta$. We simultaneously obtain the
allowed range for the $\Delta F=2$ amplitudes which can be used to
test any extension of the SM, and the prediction for the
time-dependent CP asymmetry $S_{J/\Psi \phi}$. For all our analyses we
use the method described in refs.~\cite{utfit,utnp} and the input
values listed in ref.~\cite{utfitsite}.

\begin{figure}[htb!]
\begin{center}
\includegraphics[width=0.23\textwidth]{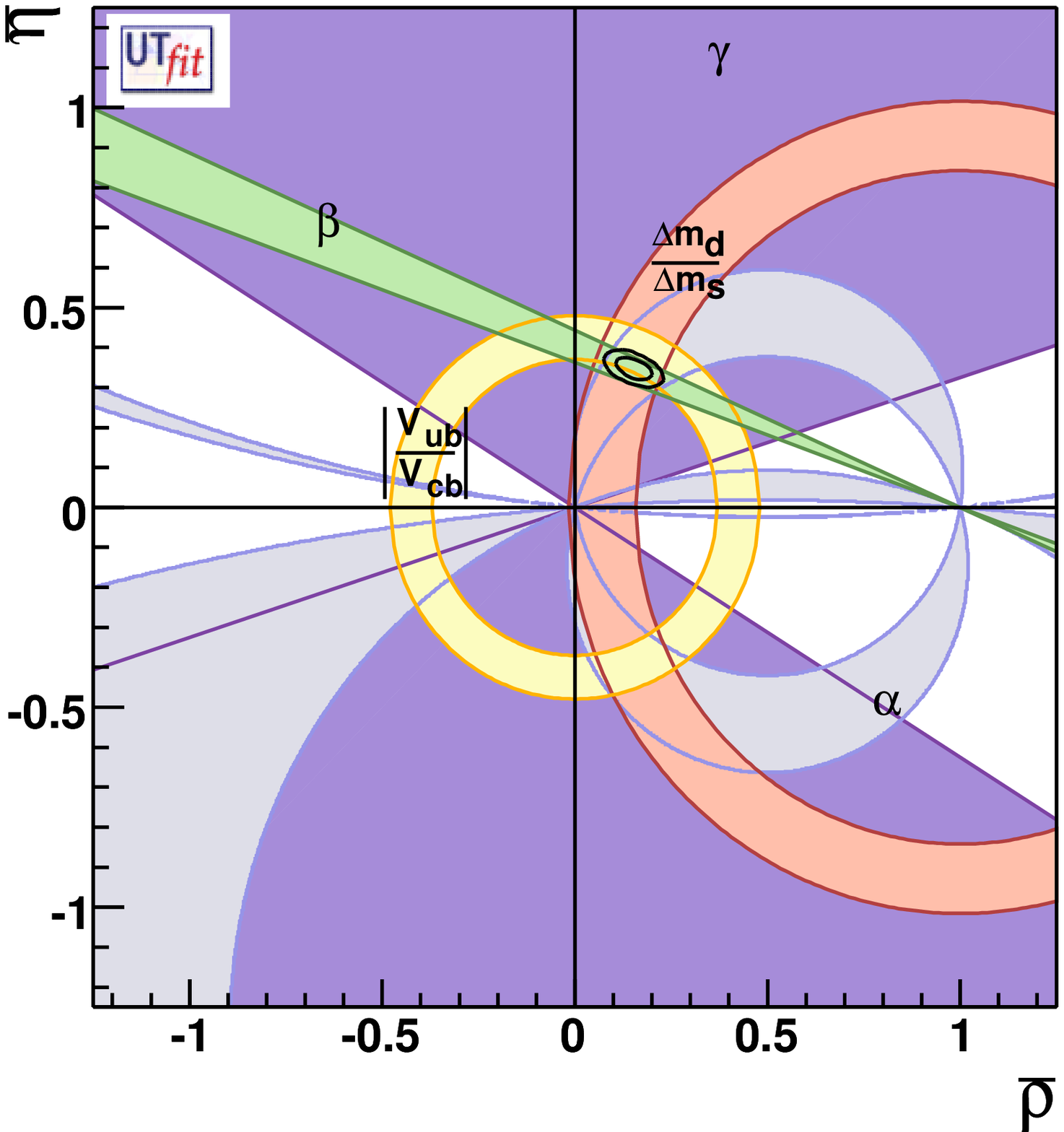}
\includegraphics[width=0.23\textwidth]{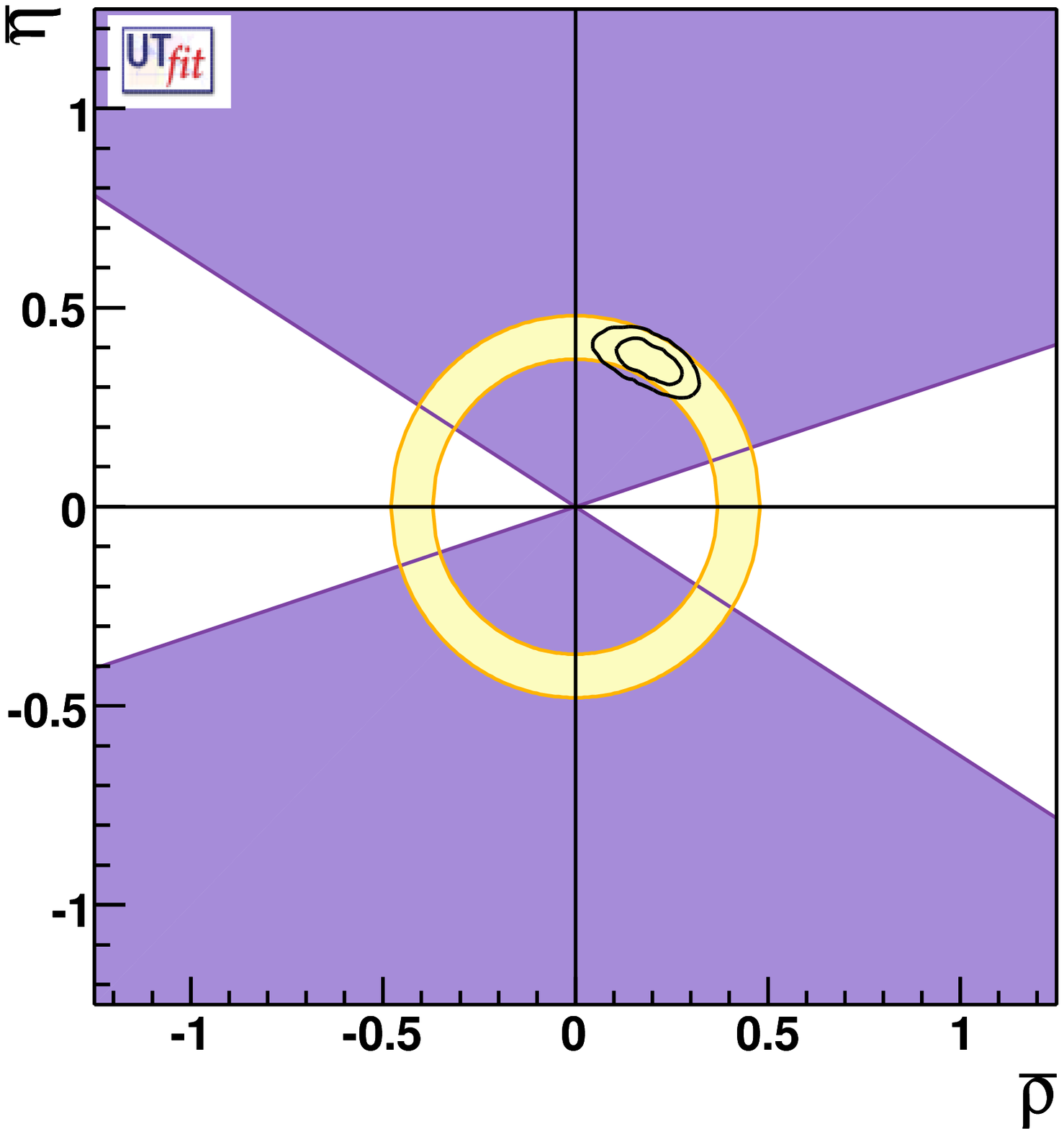}
\caption{%
  Determination of $\bar \rho$ and $\bar \eta$ from
   the constraints on $\alpha$, $\beta$, $\gamma$,
   $\vert V_{ub}/V_{cb}\vert$,
   $\Delta m_d/\Delta m_s$ (UUT fit, left) and
   from the constraints on $\alpha$, $\beta$, $\gamma$,
   $\vert V_{ub}/V_{cb}\vert$, $\Delta m_d$,
   $\Delta m_s$, $\varepsilon_K$, $A_\mathrm{SL}$, $A_\mathrm{CH}$ and
   $\Delta \Gamma_q/\Gamma_q$ (generalized NP fit, right). In the right
   plot, only tree-level constraints are shown.}
\label{fig:uut}
\end{center}
\end{figure}

In the context of MFV extensions of the SM, it is possible to
determine the parameters of the CKM matrix independently of the
presence of NP, using the UUT construction, which is independent of NP
contributions. In particular, all the constraints from tree-level
processes and from the angle measurements are valid and the NP
contribution cancels out in the $\Delta m_d/\Delta m_s$ ratio; the
only NP dependent quantities are $\varepsilon_K$ and (individually)
$\Delta m_d$ and $\Delta m_{s}$, because of the shifts $\delta S_0^K$
and $\delta S_0^B$ of the Inami-Lim functions in $K$--$\bar K$ and
$B_{d,s}$--$\bar B_{d,s}$ mixing processes. With only one Higgs
doublet or at small $\tan \beta$, these two contributions are
dominated by the Yukawa coupling of the top quark and are forced to be
equal. For large $\tan\beta$, the additional contribution from the
bottom Yukawa coupling cannot be neglected and the two quantities are
in general different. In both cases, one can use the output of the UUT
given in Tab.~\ref{tab:uut} and in the left plot of Fig.~\ref{fig:uut}
to obtain a constraint on $\delta S_0^{K,B}$ using $\varepsilon_K$ and
$\Delta m_d$.  We get $\delta S_0=\delta S_0^K=\delta S_0^B=-0.12\pm
0.32$ for small $\tan\beta$, while for large $\tan\beta$ we obtain
$\delta S_0^B=0.26\pm0.72$ and $\delta S_0^K=-0.18\pm0.38$. Using the
procedure detailed in ~\cite{gino}, these bounds can be translated
into lower bounds on the MFV scale $\Lambda$:
\begin{eqnarray}
\Lambda  &>& 
5.9 \mathrm{~TeV~@95\%~Prob.~for~small~}\tan\beta \nonumber \\
\Lambda & > & 
5.4 \mathrm{~TeV~@95\%~Prob.~for~large~}\tan\beta
\label{eq:scales}
\end{eqnarray}
significantly stronger than our previous results $\Lambda>3.6$ TeV and
$\Lambda>3.2$ TeV for small and large $\tan \beta$ respectively~\cite{utnp}.

\begin{table}[h]
\begin{center}
\begin{tabular}{@{}cccc}
\hline\hline
    Parameter               &   Output  &  Parameter & Output     \\   \hline
$\overline {\rho}$  &  $0.154\pm0.032$  &
$\overline {\eta}$  &  $0.347\pm0.018$  \\ \hline
$\alpha [^{\circ}]$ &  $91\pm5~$ & 
$\beta [^{\circ}]$  &  $22.2\pm0.9~$ \\ \hline
$\gamma [^{\circ}$] &  $66\pm5~$ & 
$\sin 2 \beta_s$    &  $0.037 \pm 0.002$ \\ \hline
$\sin 2 \beta$      &  $0.704 \pm 0.023$ &
$\rm{Im} {\lambda}_t$ [$10^{-5}$] &  $14.0\pm0.8~$ \\ \hline
$V_{ub} [10^{-3}]$  & $3.69\pm0.15$  & 
$V_{cb} [10^{-2}]$  & $4.18\pm0.07$  \\ \hline
$V_{td} [10^{-3}]$  & $8.6\pm0.3$  & 
$|V_{td}/V_{ts}|$  & $0.210\pm0.008$ \\ \hline
$R_b$  & $0.381\pm0.015$  & 
$R_t$  & $0.915\pm0.033$ \\ 
\hline\hline
\end{tabular}
\end{center}
\caption {Determination of UUT parameters
from the constraints on $\alpha$, $\beta$, $\gamma$,
$\vert V_{ub}/V_{cb}\vert$, and
$\Delta m_d/\Delta m_s$ (UUT fit).
\label{tab:uut}}
\end{table}

We now turn to the UT analysis in the presence of arbitrary NP.
Following ref.~\cite{utnp}, we incorporate general NP loop contributions in
the fit in a model independent way, parametrizing the shift induced
in the $B_q$--$\bar B_q$ mixing frequency (phase) with a parameter
$C_{B_q}$ ($\phi_{B_q}$) having expectation value of one (zero) in the
SM~\cite{cfactors}:
\begin{equation} C_{B_q}
  e^{2 i \phi_{B_q}} =\frac{\langle
    B_q|H_\mathrm{eff}^\mathrm{full}|\bar{B}_q\rangle} {\langle
    B_q|H_\mathrm{eff}^\mathrm{SM}|\bar{B}_q\rangle}=
  1 +\frac{A_q^\mathrm{NP}}{A_q^\mathrm{SM}}
    e^{2 i  \phi_q^\mathrm{NP}}\label{eq:paranp}
\end{equation}
with $q=d,s$, plus an additional parameter
  $C_{\varepsilon_K} = \mathrm{Im}\langle
    K^0|H_{\mathrm{eff}}^{\mathrm{full}}|\bar{K}^0\rangle/
  \mathrm{Im}\langle
    K^0|H_{\mathrm{eff}}^{\mathrm{SM}}|\bar{K}^0\rangle$.
As shown in refs.~\cite{previousnp,utnp}, the measurements of UT
angles strongly reduced the allowed parameter space in the $B_d$
sector. On the other hand, in previous analyses the $B_s$ sector was
completely untested in the absence of stringent experimental
constraints.  Recent experimental developments allow to improve the
bounds on NP in several ways. First, the measurement of $\Delta
m_s$~\cite{dmsCDF} and of $\Delta \Gamma_s$~\cite{dgs} provide the
first constraints on the $\phi_{B_s}$ vs. $C_{B_s}$ plane. Second, the
improved measurement of $A_{SL}$ in $B_d$ decays~\cite{BaBarASL} and
the recently measured CP asymmetry in dimuon events
($A_{CH}$)~\cite{dimuonD0} further constrain the $C_{B_q}$ and
$\phi_{B_q}$ parameters. They also strongly disfavour the solution
with $\bar \rho$ and $\bar \eta$ in the third quadrant, which now has
only $1.0\%$ probability. Finally, $\Delta \Gamma_d$~\cite{dgd}
helps in reducing further the uncertainty in $C_{B_d}$.

\begin{figure*}[htb!]
\begin{center}
\includegraphics[width=0.23\textwidth]{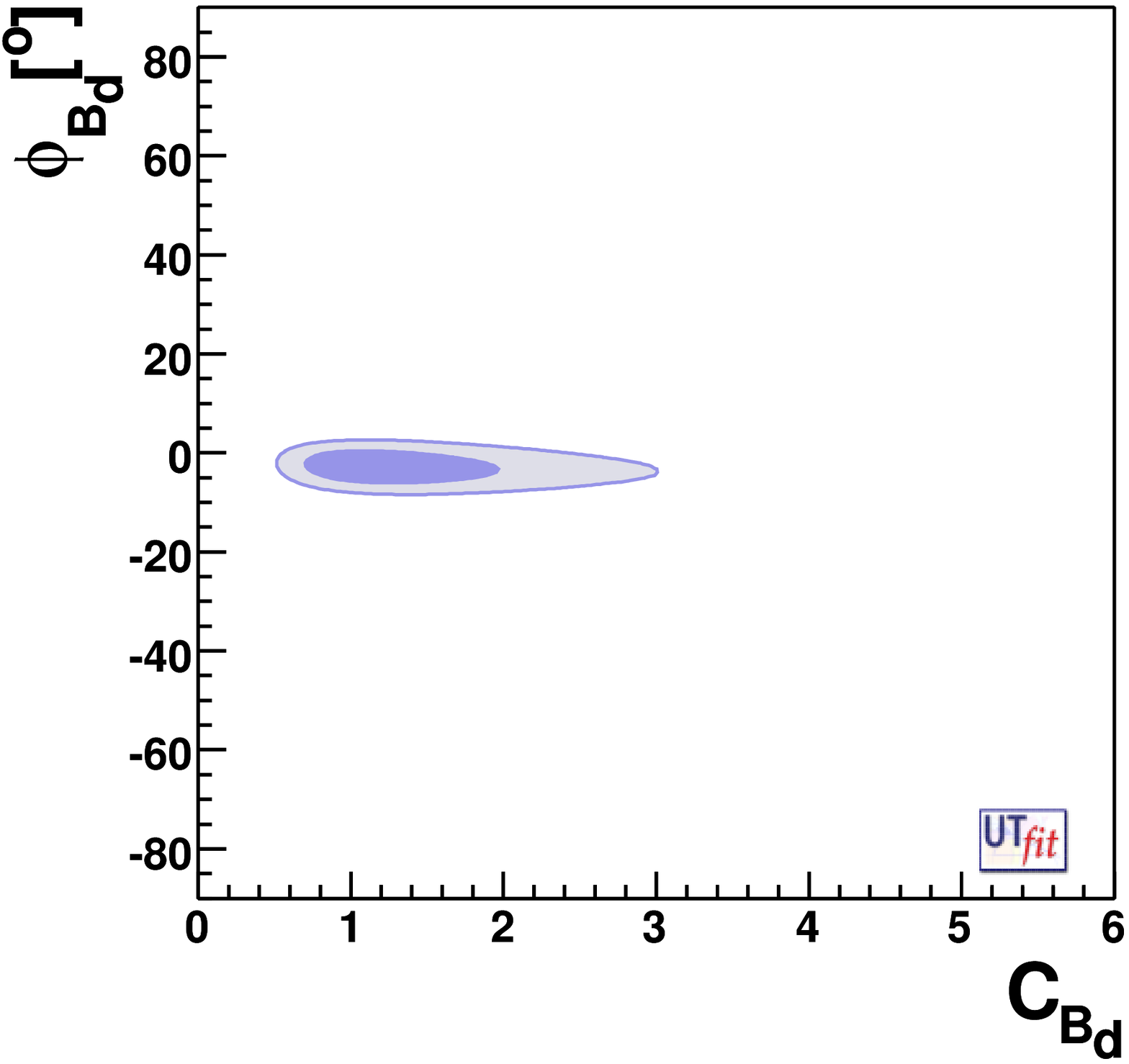}
\includegraphics[width=0.23\textwidth]{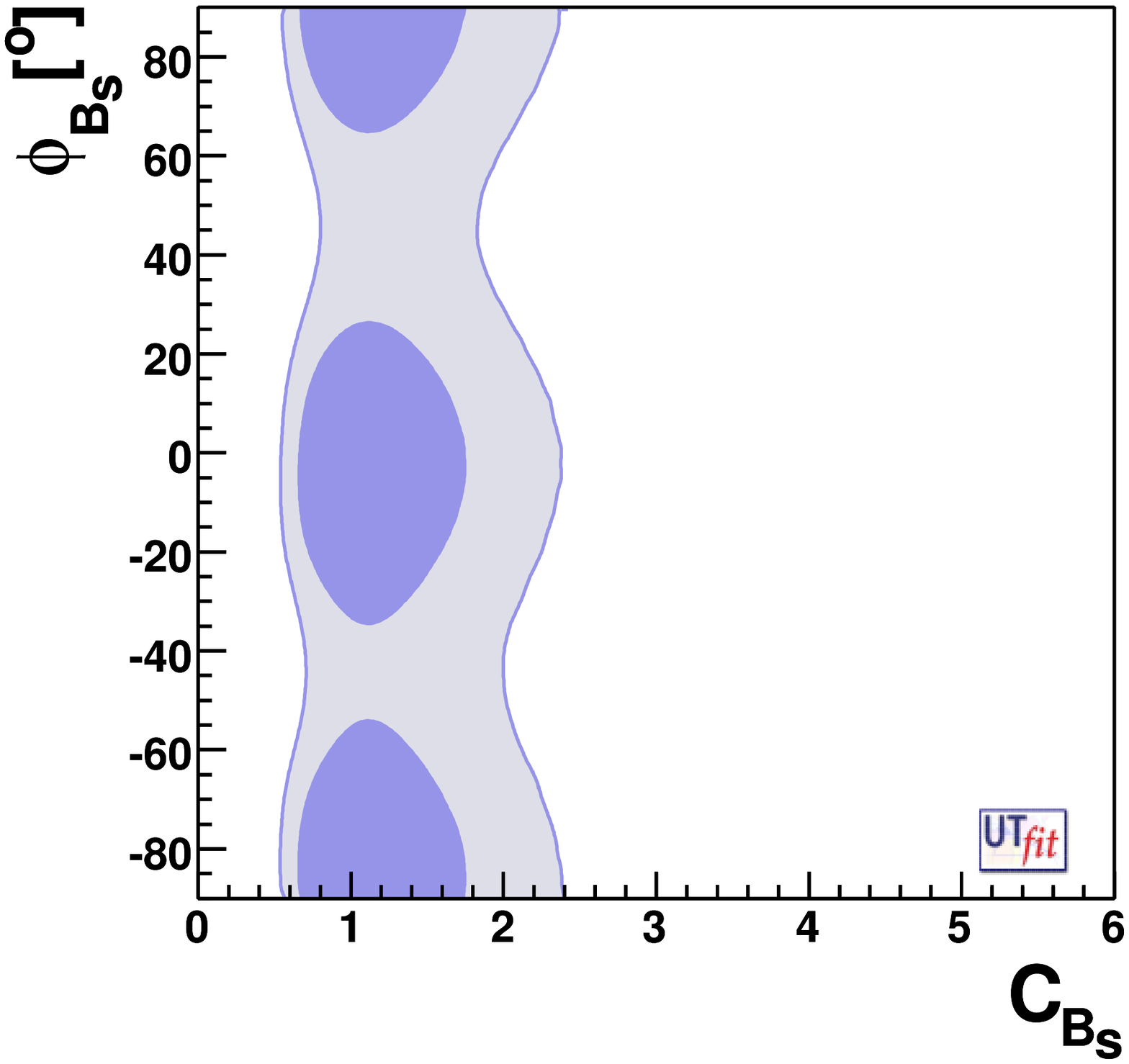}
\includegraphics[width=0.23\textwidth]{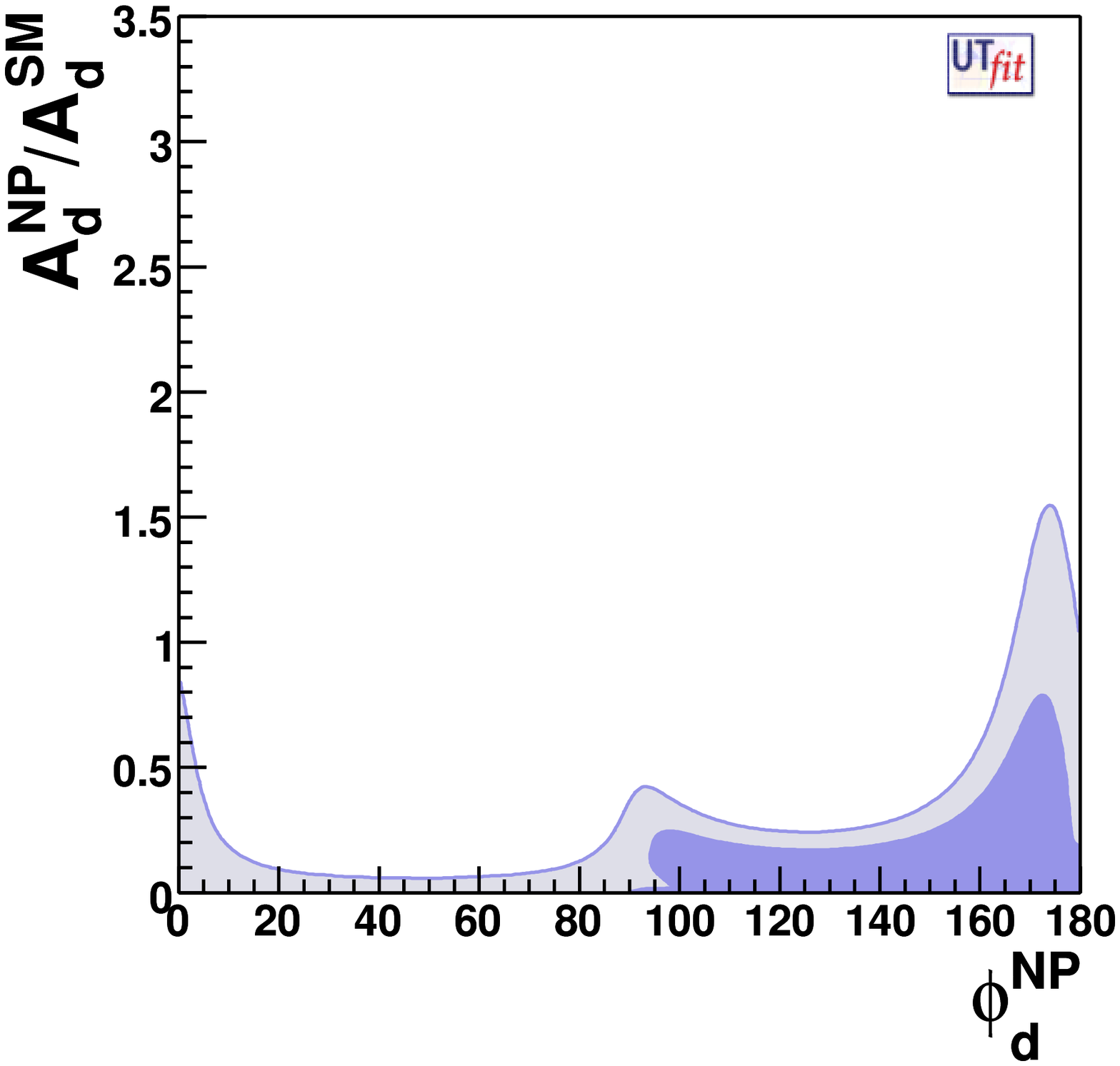}
\includegraphics[width=0.23\textwidth]{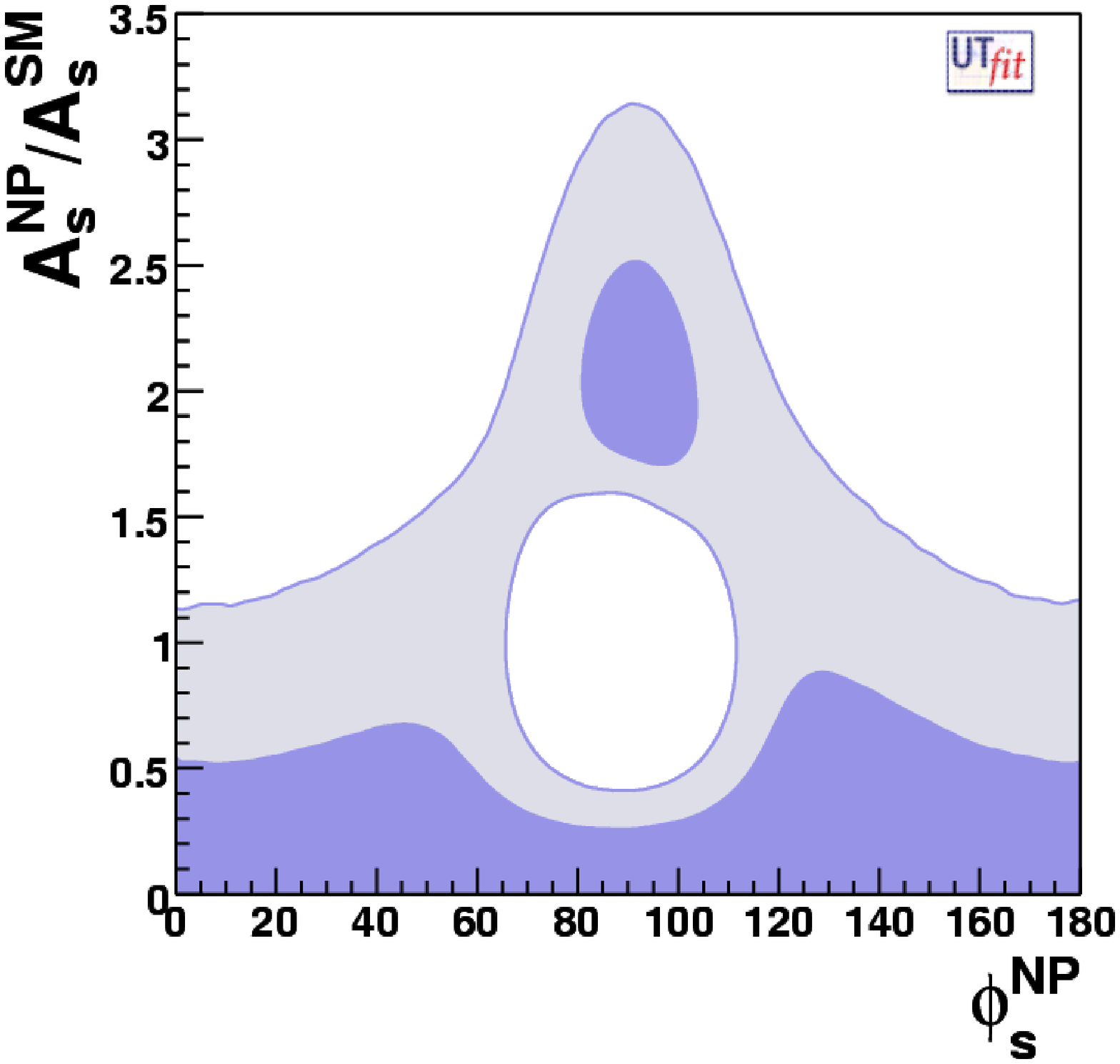}
\caption{%
  From left to right, constraints on $\phi_{B_d}$ vs. $C_{B_d}$,
  $\phi_{B_s}$ vs. $C_{B_s}$, $\phi_d^\mathrm{NP}$
  vs. $A_d^\mathrm{NP}/A_d^\mathrm{SM}$ and $\phi_s^\mathrm{NP}$
  vs. $A_s^\mathrm{NP}/A_s^\mathrm{SM}$ from the NP generalized
  analysis.}
\label{fig:npCphi}
\end{center}
\end{figure*}
\begin{figure}[htb!]
\begin{center}
\includegraphics[width=0.23\textwidth]{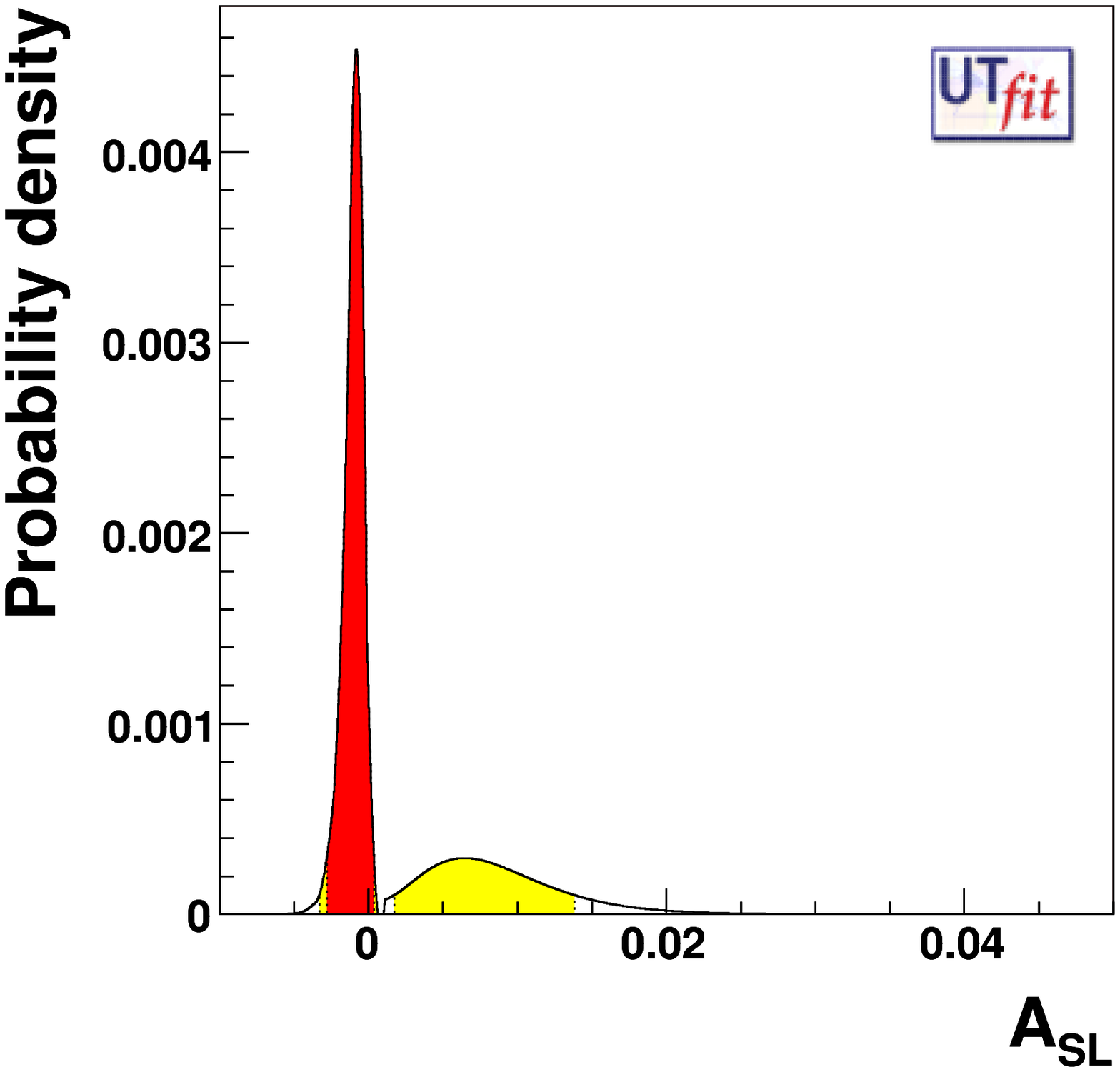} 
\includegraphics[width=0.23\textwidth]{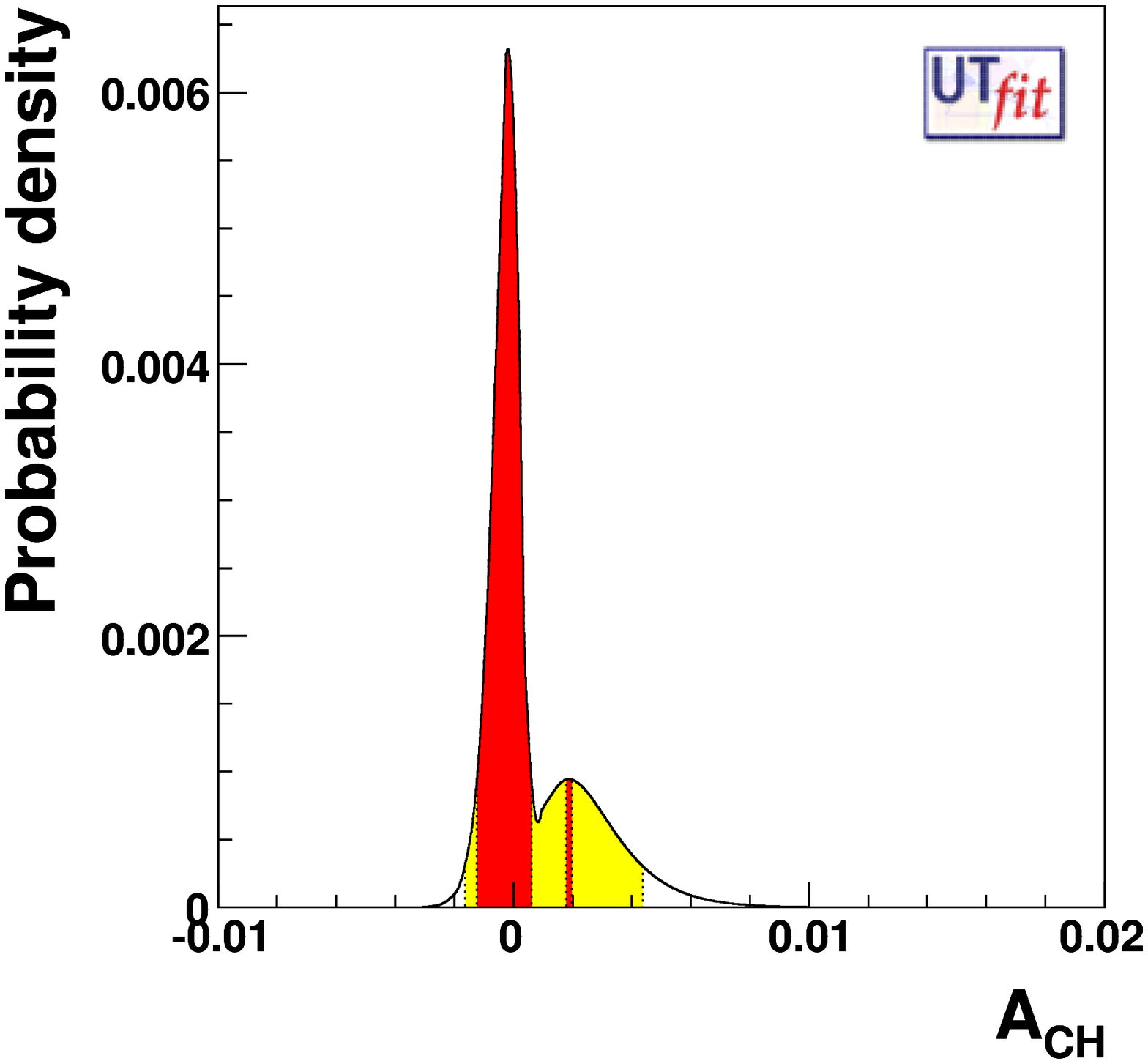} 
\caption{%
  Predictions for $A_\mathrm{SL}$ and $A_\mathrm{CH}$ in the presence
  of NP, obtained without including these observables in the fit. The
  lower peak in the p.d.f.'s correspond to values of $\rho$ and
  $\eta$ in the third quadrant.}
\label{fig:asy}
\end{center}
\end{figure}

The use of $A_{CH}$ and $\Delta \Gamma_q$ to bound $C_{B_q}$ and
$\phi_{B_q}$ deserve some explanation, while for all the other
constraints we refer the reader to ref.~\cite{utnp}. The dimuon
charge asymmetry $A_{CH}$ can be written as:
$$
\frac{(\chi - \bar
    \chi)(P_1-P_3+0.3\,P_8^\prime)}{\xi (P_1+P_3)+(1-\xi)P_2+0.28 P_7+0.5
    P_8^\prime+0.69 P_{13}}
$$
in the notation of ref.~\cite{dimuonD0}, where the definition and the
measured values for the $P$ parameters can be found. We have
$\chi=f_d \chi_d + f_s \chi_s$, $\bar \chi=f_d \bar \chi_d +f_s \bar \chi_s$ and $\xi=\chi+\bar \chi -2 \chi \bar \chi$,
where we have assumed equal semileptonic widths for $B_s$ and $B_d$
mesons, $f_d=0.397 \pm 0.010$ and  $f_s=0.107 \pm 0.011$ are the
production fractions of $B_d$ and $B_s$ mesons respectively \cite{pdg}
and $\chi_q$ and $\bar \chi_q$ are given by the expression
\begin{equation}
  \label{eq:chiq}
  \stackrel{\scriptscriptstyle{(-)}}{\chi_q}=\frac{\frac{\Delta\Gamma_q}{\Gamma_q}^2+4\frac{\Delta
      m_q}{\Gamma_q}^2}{\frac{\Delta\Gamma_q}{\Gamma_q}^2(\stackrel{\scriptscriptstyle{(-)}}{z_q}-1)
    +  
    4(2 \stackrel{\scriptscriptstyle{(-)}}{z_q}+\frac{\Delta
      m_q}{\Gamma_q}^2(1+
    \stackrel{\scriptscriptstyle{(-)}}{z_q}))}
\end{equation}
with $z_q=\vert q/p \vert_q^2$ and
$\bar z_q=\vert p/q \vert_q^2$.
Finally, using the results of~\cite{tarantino} and following the
notation of~\cite{utnp}, we have
\begin{equation}
\frac{\Delta \Gamma_q}{\Delta m_q} = \mathrm{Re}\, \mathcal{P}\,,\qquad
\left\vert\frac{q}{p}\right\vert_q-1=-\frac{1}{2}\mathrm{Im}\, \mathcal{P}
\end{equation}
where
%\begin{widetext}
  \begin{eqnarray}
   &&\!\!\!\! \mathcal{P}  = -2\frac{\kappa}{C_{B_q}} \left\{e^{2
        \phi_{B_q}} \left(n_1+\frac{n_6 B_2+n_{11}}{B_1}\right)
        \right.\nonumber \\
    &&\!\!\!\!  -\frac{e^{(\phi_{q}^\mathrm{SM}+2 \phi_{B_q})}}{R_t^q}
      \left(n_2+\frac{n_7
          B_2+n_{12}}{B_1}\right) 
    +\frac{e^{2 (\phi_{q}^\mathrm{SM}+\phi_{B_q})}}{R_t^{q^2}}\nonumber \\
   &&\!\!\!\! \left(n_3+\frac{n_8 B_2 + n_{13}}{B_1}\right)
    +e^{(\phi_q^\mathrm{Pen}+2 \phi_{B_q})} C_q^\mathrm{Pen}
    \left(n_4+n_9 \frac{ B_2}{B_1}\right)\nonumber \\
    &&\!\!\!\! \left. -e^{(\phi_{q}^\mathrm{SM}+\phi_q^\mathrm{Pen}+2 \phi_{B_q})}
      \frac{C_q^\mathrm{Pen}}{R_t^q}
      \left(n_5+n_{10}\frac{B_2}{B_1}\right)\right\} \label{eq:dgammafull}
  \end{eqnarray}
%\end{widetext}
with $\kappa=-2 \pi m_b^2/(3 M_W^2 \eta_B S_0(x_t))$, the $B$
parameters and the magic numbers $n_i$ given in ref.~\cite{utnp}
($SU(3)$ breaking effects in the magic numbers can be neglected given
the present errors) and $R_t^q=\vert V_{tq}V_{tb}^*\vert/\vert
V_{cq}V_{cb}^*\vert$. As discussed in ref.~\cite{utnp},
$C_q^\mathrm{Pen}$ and $\phi_q^\mathrm{Pen}$ parametrize possible NP
contributions to $\Delta B=1$ penguins. Concerning $\Delta \Gamma_s$,
since the available experimental measurements are not directly
sensitive to the phase of the mixing amplitude, they are actually a
measurement of $\Delta \Gamma_s \cos 2 (\phi_{B_s}-\beta_s)$ in the
presence of NP \cite{dgammanp}. To assess the constraining power of
leptonic asymmetries and width differences, we compare the SM
predictions and the experimental results with the predictions in the
presence of NP, see Tab.~\ref{tab:aedg} and Fig.~\ref{fig:asy}.
\begin{table}[htb!]
  \centering
  \begin{tabular}{ccccc}
  \hline \hline
   & SM & SM+NP & exp & ref\\ \hline
    $10^{3} A_\mathrm{SL}$& $-0.71 \pm 0.12$ & see
    Fig.~\ref{fig:asy} &  
    $-0.3 \pm 5$ &\cite{BaBarASL}\\\hline 
    $10^{3} A_\mathrm{CH}$&$-0.23 \pm 0.05$& see
    Fig.~\ref{fig:asy} &
    $-13 \pm 12 \pm 8$ &\cite{dimuonD0} \\\hline 
    $10^{3}\Delta \Gamma_d/\Gamma_d$ & $3.3 \pm 1.9$ & $2.0 \pm
    1.8$ & 
    $9 \pm 37$ &\cite{dgd}\\\hline 
    $\Delta \Gamma_s/\Gamma_s$ & $0.10 \pm 0.06$ & $0.00 \pm 0.08$ &
    $0.25 \pm 0.09$ &\cite{dgs}\\\hline 
    \hline
\end{tabular}
\caption{Predictions for
  $A_\mathrm{SL}$, $A_\mathrm{CH}$ and $\Delta \Gamma_q/\Gamma_q$
  in the SM or in the presence of NP, obtained without including 
  these observables in the fit.}
\label{tab:aedg}
\end{table}
We see that NP can produce dilepton asymmetries ($\Delta \Gamma_s$)
much larger (smaller) than the SM, so that including them in the fit
improves the constraints on NP. For each value of ($C_{B_q}$,
$\phi_{B_q}$) we compute $A_\mathrm{SL}$, $A_\mathrm{CH}$ and $\Delta
\Gamma_s \cos 2 (\phi_{B_s}-\beta_s)$ and use the experimental values
to compute the weight of the given configuration. In ref.~\cite{nir}, the measurement of $A_\mathrm{CH}$ was used
in a different way:  $A_\mathrm{CH}$ was combined with the experimental value of $A_\mathrm{SL}$ to obtain a value for $A_\mathrm{SL}^s$. In principle, our method takes into account
the correlations between the theoretical predictions for $A_\mathrm{SL}$ and $A_\mathrm{SL}^s$. In addition, using $A_\mathrm{CH}$ instead of $A_\mathrm{SL}^s$ is more constraining since the theoretical range for
$A_\mathrm{SL}$ is smaller than the present experimental
error. In practice, however, these two effects are rather
small.

\begin{figure}[tb!]
\begin{center}
\includegraphics[width=0.23\textwidth]{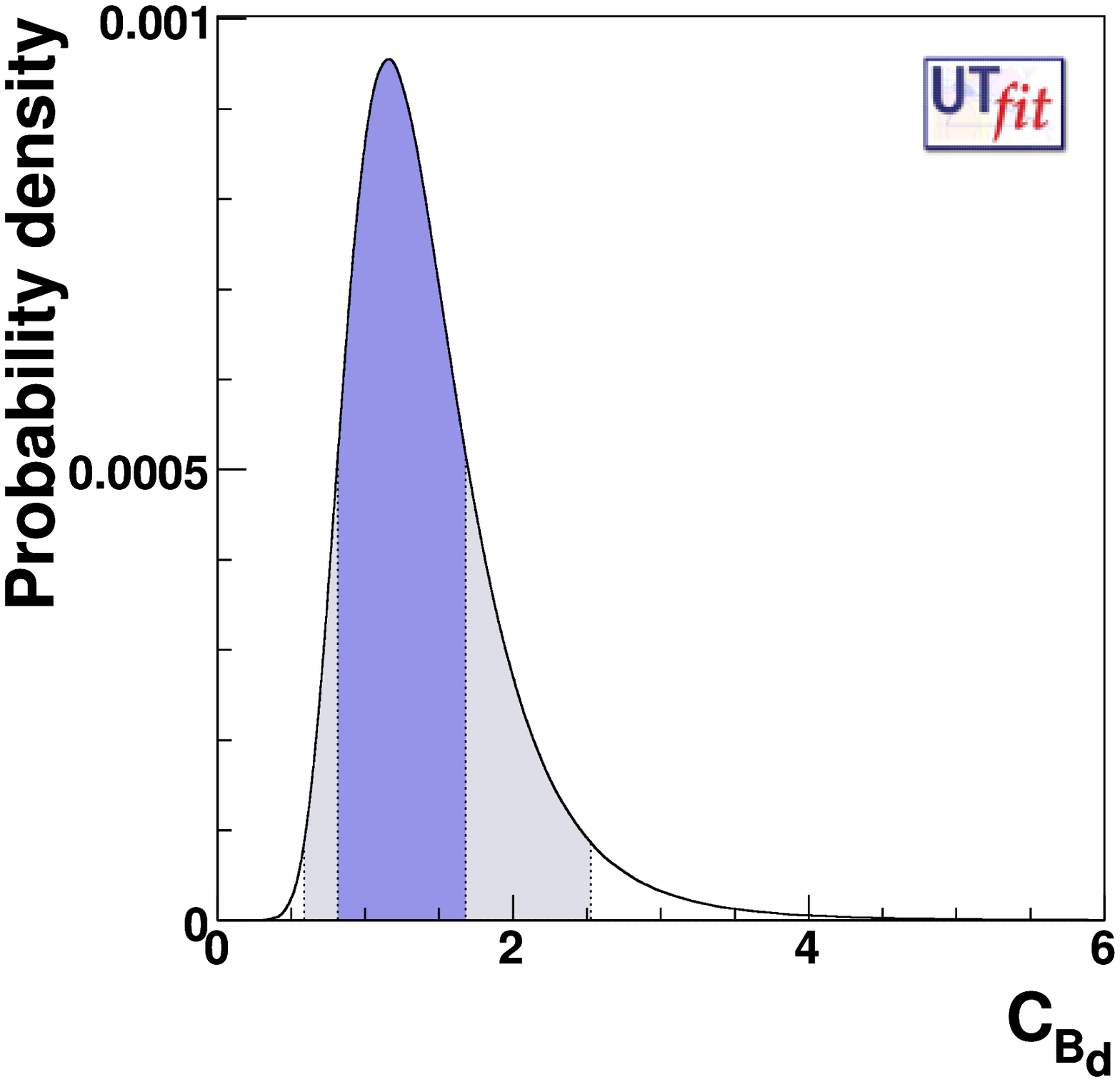}
\includegraphics[width=0.23\textwidth]{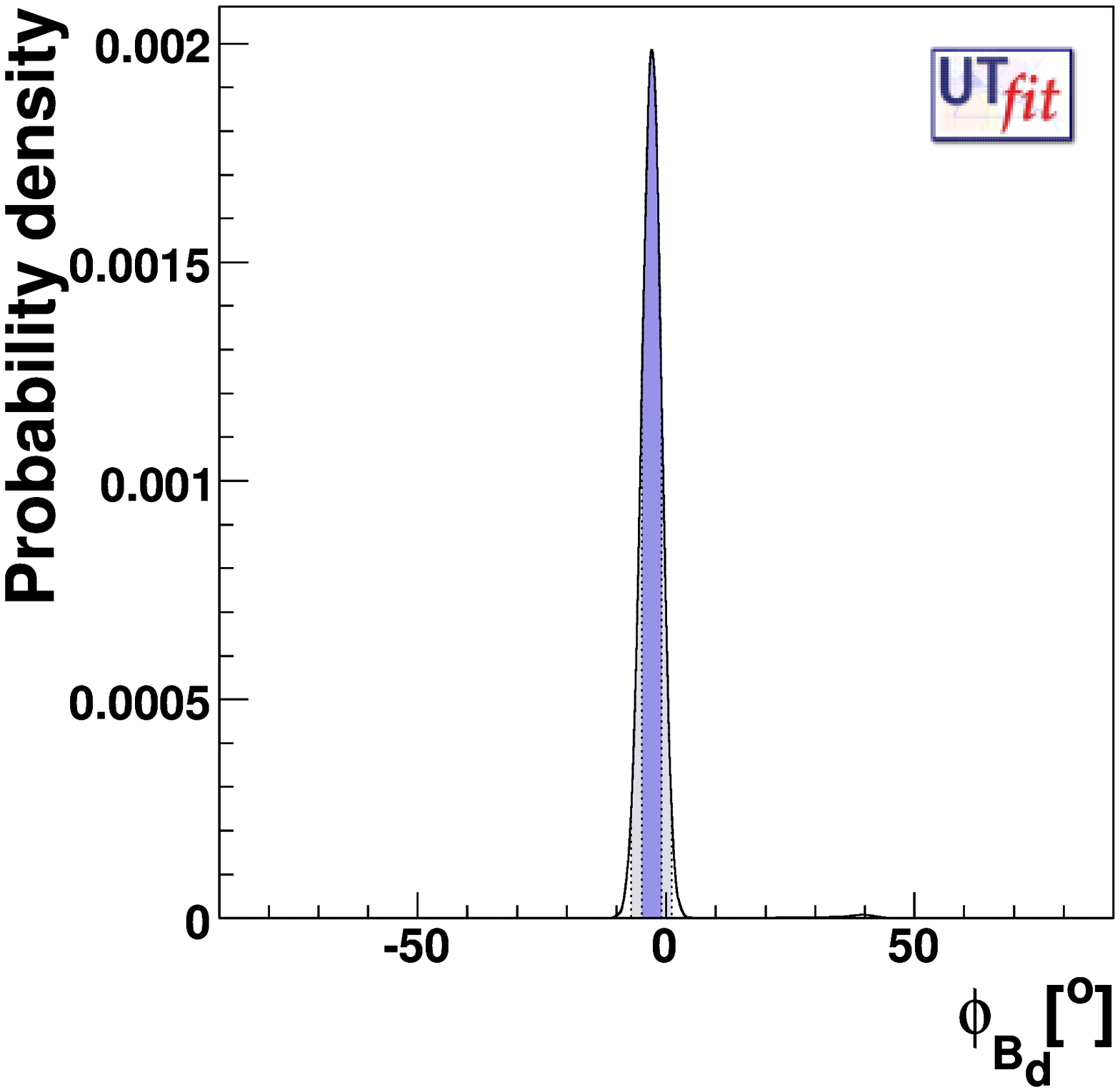}\\
\includegraphics[width=0.23\textwidth]{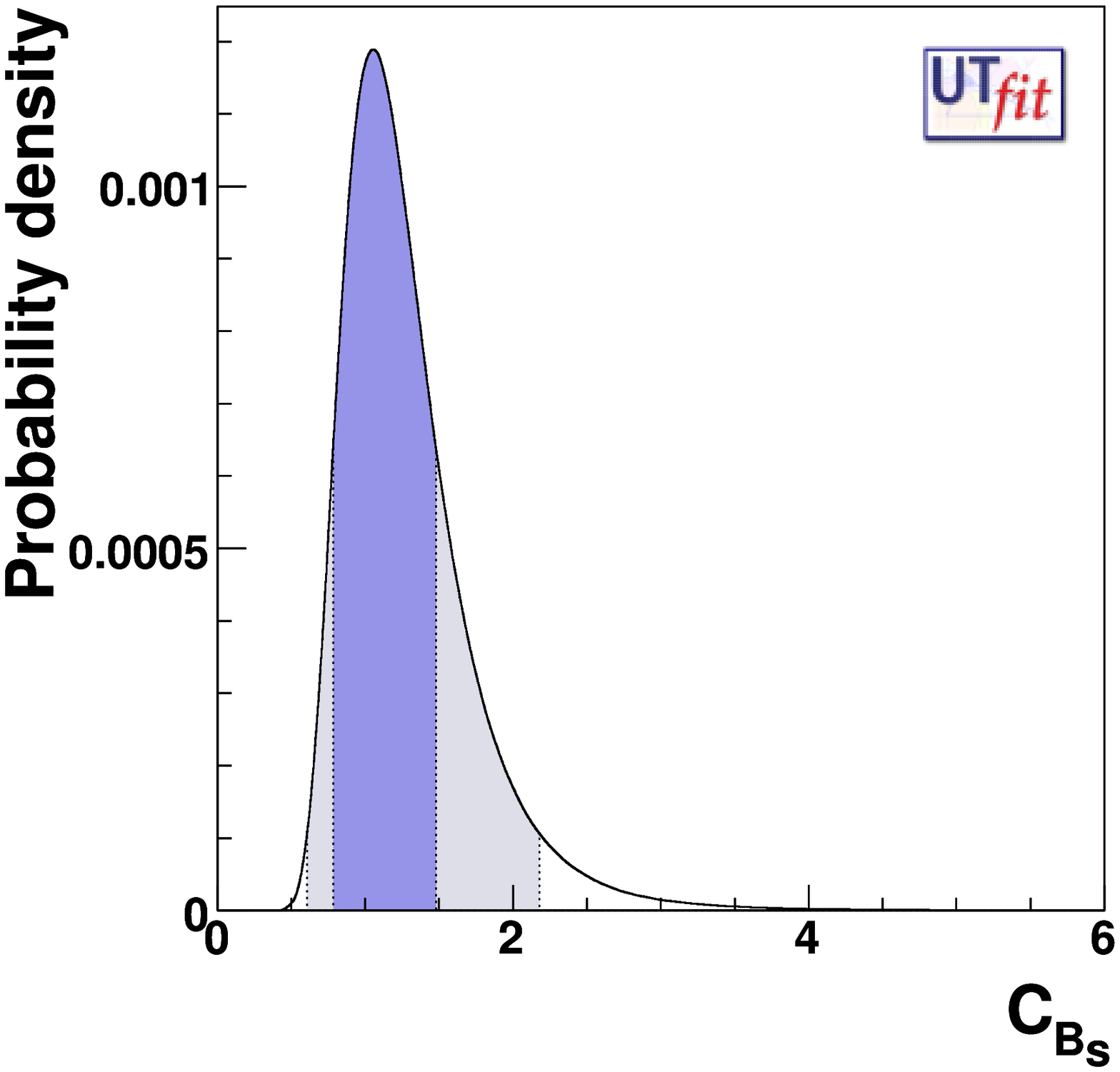}
\includegraphics[width=0.23\textwidth]{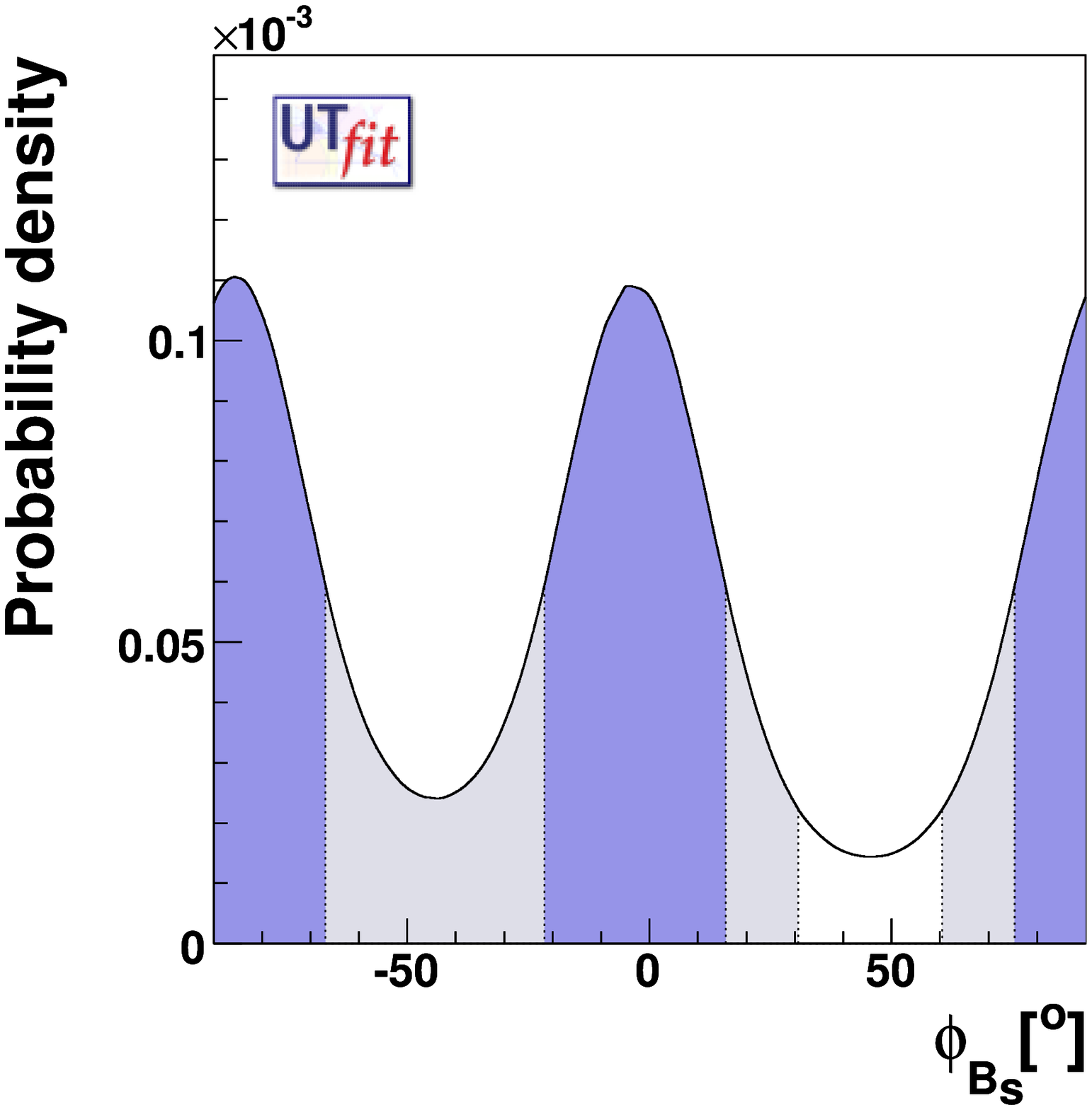}\\
\includegraphics[width=0.23\textwidth]{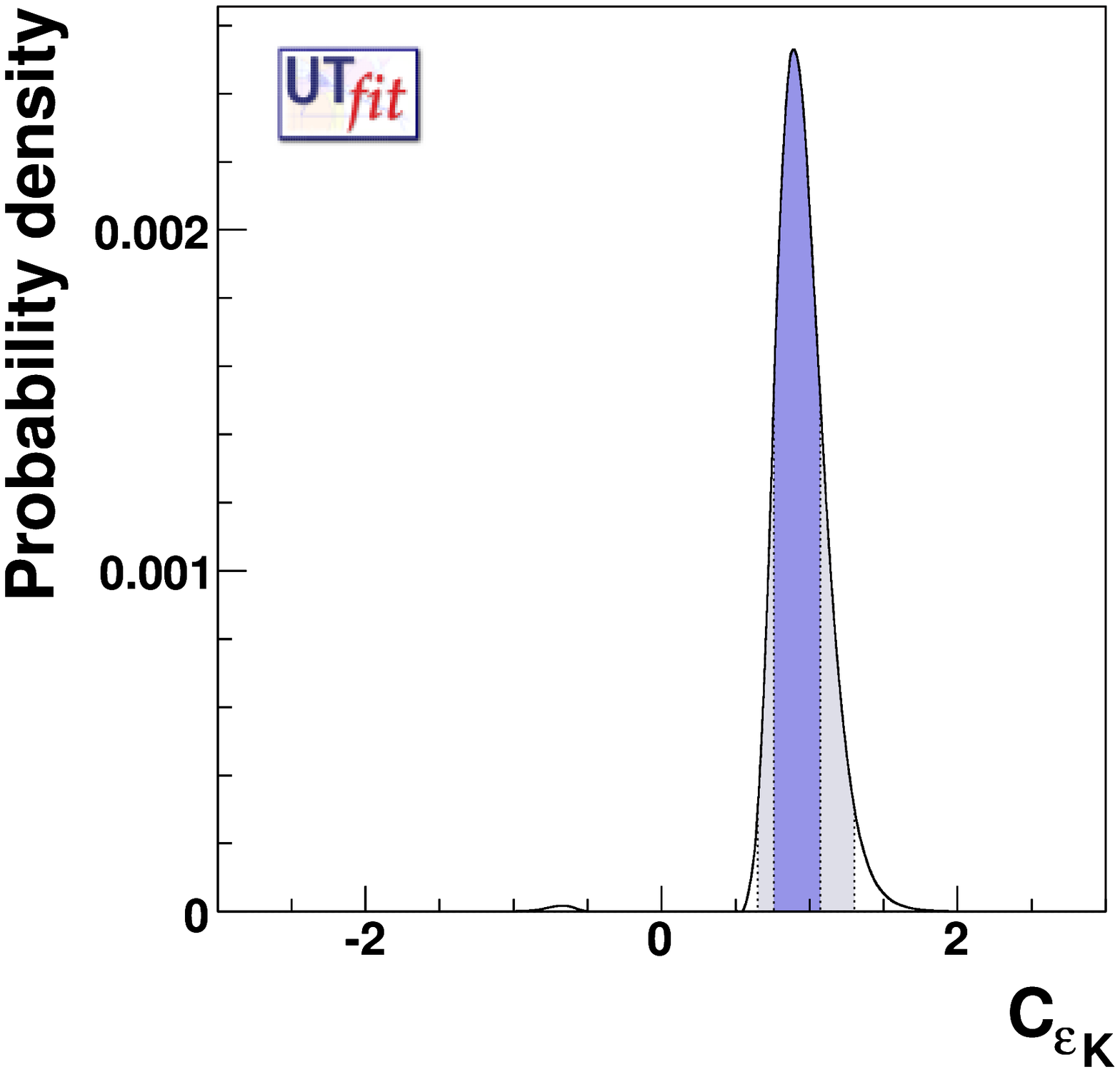}
\includegraphics[width=0.23\textwidth]{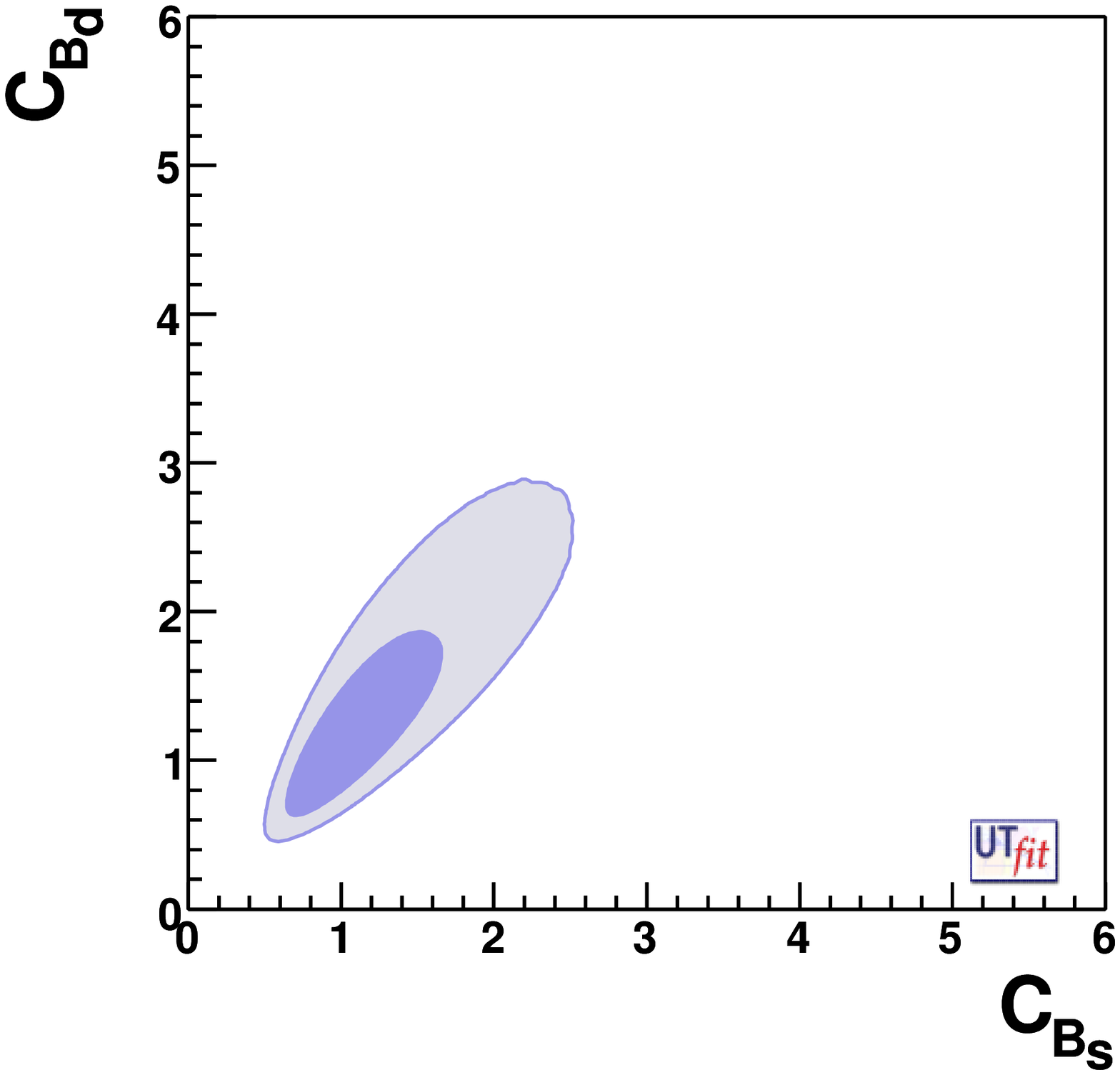}
\caption{%
  Constraints on $\phi_{B_q}$, $C_{B_q}$ and
  $C_{\varepsilon_K}$  coming from the NP generalized analysis.
  The correlation between $C_{B_d}$ and $C_{B_s}$ is also shown.}
\label{fig:1Dnp}
\end{center}
\end{figure}

The result of the fit is summarized in Tab.~\ref{tab:NP}. The bound on
$\bar\rho$ and $\bar\eta$ is also shown in right plot of Fig.~\ref{fig:uut}, while the
bounds on the two $\phi_B$ vs. $C_B$ planes are given in
Fig.~\ref{fig:npCphi}, together with the corresponding regions in the
$\phi_q^\mathrm{NP}$ vs. $A_q^\mathrm{NP}/A_q^\mathrm{SM}$ planes. The
distributions for $C_{B_q}$, $\phi_{B_q}$ and $C_{\varepsilon_K}$ are
shown in Fig.~\ref{fig:1Dnp}.  We see that the \emph{non-standard}
solution for the UT with its vertex in the third quadrant, which was
present in the previous analysis~\cite{utnp}, is now absent thanks to
the improved value of $A_\mathrm{SL}$ by the BaBar Collaboration and
to the measurement of $A_\mathrm{CH}$ by the D0 Collaboration (the
lower peaks in Fig.~\ref{fig:asy} correspond to the
\emph{non-standard} solution and are now excluded). Furthermore, the
measurement of $\Delta m_s$ strongly constrains $C_{B_s}$, so that
$C_{B_s}$ is already known better than $C_{B_d}$. Finally, $A_{CH}$
and $\Delta \Gamma_s$ provide stringent constraints on
$\phi_{B_s}$. Taking these constraints into account, we obtain
\begin{equation}
  \label{eq:sjpsiphi}
  S_{J/\Psi \phi}=0.09 \pm 0.60\,,
\end{equation}
leaving open the possibility of observing large values of $S_{J/\Psi
  \phi}$ at LHCb. We point out an interesting correlation between the
values of $C_{B_d}$ and $C_{B_s}$ that can be seen in
Fig.~\ref{fig:1Dnp}. This completely general correlation is present
since lattice QCD determines quite precisely the ratio $\xi^2$ of the
matrix elements entering $B_s$ and $B_d$ mixing amplitudes
respectively.

We conclude by noting that the fit produces a nonzero central value of
$\phi_{B_d}$. This is due to the difference in the SM fit between the
angles measurement (in particular $\sin 2 \beta$) and the sides
measurement (in particular $V_{ub}$ inclusive). More details on this
difference can be found in ref.~\cite{instantSM}. Further
improvements in experimental data and in theoretical analyses are
needed to tell whether this is just a fluctuation or we are really
seeing a first hint of NP in the flavour sector.

\begin{table}[h]
\begin{center}
\begin{tabular}{@{}cccc}
\hline\hline
    Parameter               &   Output  &  Parameter & Output     \\   \hline
\hline
$C_{B_d}$ & $1.25 \pm 0.43$ &
$\phi_{B_d} [^{\circ}]$ & $-2.9 \pm 2.0$ \\ 
\hline
$C_{B_s}$ & $1.13 \pm 0.35$ &
$\phi_{B_s} [^{\circ}]$ & $(-3 \pm 19) \cup (94 \pm 19)$ \\ 
\hline
$C_{\varepsilon_K}$ & $0.92 \pm 0.16$ & 
%\multicolumn{2}{c}{$[0.64,1.47]$ @ $95$\% prob.} 
\\ \hline
\hline
$\overline {\rho}$  &  $0.20\pm0.06$  &
$\overline {\eta}$  &  $0.36\pm0.04$  \\ \hline
$\alpha [^{\circ}]$ &  $93\pm 9$ &
$\beta [^{\circ}]$ &  $24\pm 2$ \\ \hline
$\gamma[^{\circ}]$ &  $62\pm9~$ &
$\rm{Im} {\lambda}_t$[$10^{-5}$] &  $14.6\pm 1.4$ \\ \hline
$V_{ub}[10^{-3}]$  & $4.01\pm 0.25$ &
$V_{cb}[10^{-2}]$  & $4.15\pm 0.07$  \\ \hline
$V_{td}[10^{-3}]$  & $8.33\pm 0.61$ &
$|V_{td}/V_{ts}|$  & $0.203\pm 0.015$  \\ \hline
$R_b$  & $0.416\pm 0.027$ &
$R_t$  & $0.887\pm 0.063$ \\ \hline
$\sin2\beta$ & $0.748 \pm 0.040$ &
$\sin2\beta_s$ & $0.039 \pm 0.004$ \\ \hline
\hline
\end{tabular}
\end{center}
\caption {Determination of UT and NP parameters
from the NP generalized fit.}
\label{tab:NP}
\end{table}

We thank A.J. Buras, S. Giagu, A. Lenz, M. Rescigno and A. Weiler for useful discussions. This work has been supported in part by the EU network ``The quest for unification'' under the contract MRTN-CT-2004-503369.

\end{document}